\documentclass[runningheads]{llncs}
\usepackage{graphicx}
\usepackage{comment}
\usepackage{amsmath,amssymb} 
\usepackage{color}

\usepackage{cite}
\usepackage{multirow}
\usepackage{booktabs}
\usepackage[table]{xcolor}

\usepackage{subcaption}
\usepackage{float}

\usepackage[width=122mm,left=12mm,paperwidth=146mm,height=193mm,top=12mm,paperheight=217mm]{geometry}

\begin{document}
\pagestyle{headings}
\mainmatter
\def\ECCVSubNumber{1003}  

\title{Defocus Deblurring Using Dual-Pixel Data} 

\titlerunning{Defocus Deblurring Using Dual-Pixel Data}
%
\author{Abdullah Abuolaim\inst{1} \and
Michael S. Brown\inst{1,2}}
\authorrunning{A. Abuolaim et al.}
%
\institute{York University, Toronto, Canada \and
Samsung AI Center, Toronto, Canada \\
\email{\{abuolaim,mbrown\}@eecs.yorku.ca}}
\maketitle

\begin{abstract}
Defocus blur arises in images that are captured with a shallow depth of field due to the use of a wide aperture.  Correcting defocus blur is challenging because the blur is spatially varying and difficult to estimate.  We propose an effective defocus deblurring method that exploits data available on dual-pixel (DP) sensors found on most modern cameras. DP sensors are used to assist a camera's auto-focus by capturing two sub-aperture views of the scene in a single image shot.  The two sub-aperture images are used to calculate the appropriate lens position to focus on a particular scene region and are discarded afterwards.  We introduce a deep neural network (DNN) architecture that uses these discarded sub-aperture images to reduce defocus blur. A key contribution of our effort is a carefully captured dataset of 500 scenes (2000 images) where each scene has: (i) an image with defocus blur captured at a large aperture; (ii) the two associated DP sub-aperture views; and (iii) the corresponding all-in-focus image captured with a small aperture.  Our proposed DNN produces results that are significantly better than conventional single image methods in terms of both quantitative and perceptual metrics -- all from data that is already available on the camera but ignored. The dataset, code, and trained models are available at \url{https://github.com/Abdullah-Abuolaim/defocus-deblurring-dual-pixel}.

\end{abstract}

\section{Introduction}

\begin{figure}[t]
\includegraphics[width=\linewidth]{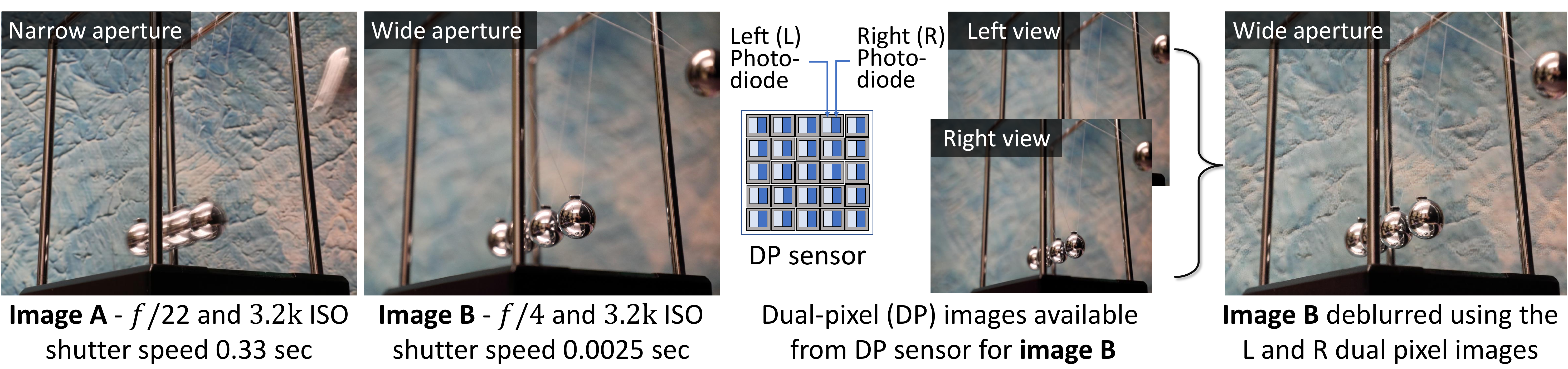}
\caption{Images A and B are of the same scene and same approximate exposure. Image A is captured with a narrow aperture (f/22) and slow shutter speed.  Image A has a wide depth of field (DoF) and little defocus blur, but exhibits motion blur from the moving object due to the long shutter speed.  Image B is captured with a wide aperture (f/4) and a fast shutter speed. Image B exhibits defocus blur due to the shallow DoF, but has no motion blur.  Our proposed DNN uses the two sub-aperture views from the dual-pixel sensor of image B to deblur image B, resulting in a much sharper image.}\label{fig:teaser}
\end{figure}

This paper addresses the problem of defocus blur.  To understand why defocus blur is difficult to avoid, it is important to understand the mechanism governing image exposure.  An image's exposure to light is controlled by adjusting two parameters: shutter speed and aperture size.  The shutter speed controls the duration of light falling on the sensor, while the aperture controls the amount of light passing through the lens.  The reciprocity between these two parameters allows the same exposure to occur by fixing one parameter and adjusting the other.  For example, when a camera is placed in {\it aperture-priority} mode, the aperture remains fixed while the shutter speed is adjusted to control how long light is allowed to pass through the lens.   The drawback is that a slow shutter speed can result in motion blur if the camera and/or an object in the scene moves while the shutter is open, as shown in Fig.~\ref{fig:teaser}.  Conversely, in {\it shutter-priority} mode, the shutter speed remains fixed while the aperture adjusts its size.  The drawback of a variable aperture is that a wide aperture results in a shallow depth of field (DoF), causing defocus blur to occur in scene regions outside the DoF, as shown in Fig.~\ref{fig:teaser}.  There are many computer vision applications that require a wide aperture but still want an all-in-focus image.  An excellent example is cameras on self-driving cars, or cameras on cars that map environments, where the camera must use a fixed shutter speed and the only way to get sufficient light is a wide aperture at the cost of defocus blur.

Our aim is to reduce the unwanted defocus blur.  The novelty of our approach lies in the use of data available from dual-pixel (DP) sensors used by modern cameras. DP sensors are designed with two photodiodes at each pixel location on the sensor. The DP design provides the functionality of a simple two-sample light-field camera and was developed to improve how cameras perform autofocus.  Specifically, the two-sample light-field provides two sub-aperture views of the scene, denoted in this paper as {\it left} and {\it right} views.  The light rays coming from scene points that are within the camera's DoF (i.e., points that are in focus) will have no difference in phase between the left and right views.  However, light rays coming from scene points outside the camera's DoF (i.e., points that are out of focus) will exhibit a detectable disparity in the left/right views that is directly correlated to the amount of defocus blur. We refere to it as {\it defocus disparity}. Cameras use this phase shift information to determine how to move the lens to focus on a particular location in the scene.  After autofocus calculations are performed, the DP information is discarded by the camera's hardware.

\noindent{\textbf{Contribution.}}~We propose a deep neural network (DNN) to perform defocus deblurring that uses the DP images from the sensor available at capture time.  In order to train the proposed DNN, a new dataset of 500 carefully captured images exhibiting defocus blur and their corresponding all-in-focus image is collected.  This dataset consists of 2000 images -- 500 DoF blurred images with their 1000 DP sub-aperture views and 500 corresponding all-in-focus images -- all at full-frame resolution (i.e., $6720\times4480$ pixels).  Using this training data, we propose a DNN architecture that is trained in an end-to-end manner to directly estimate a sharp image from the left/right DP views of the defocused input image. Our approach is evaluated against conventional methods that use only a single input image and show that our approach outperforms the existing state-of-the-art approaches in both signal processing and perceptual metrics.  Most importantly, the proposed method works by using the DP sensor images that are a free by-product of modern image capture.

\section{Related work}

Related work is discussed regarding (1) defocus blur, (2) datasets, and (3) applications exploiting DP sensors.

\noindent{\textbf{Defocus deblurring.}}~Related methods in the literature can be categorized into: (1) defocus detection methods~\cite{golestaneh2017spatially,shi2014discriminative,tang2019defusionnet,yi2016lbp,zhao2018defocus,zhao2019enhancing} or (2) defocus map estimation and deblurring methods~\cite{d2016non,karaali2017edge,lee2019deep,park2017unified,shi2015just}. While defocus detection is relevant to our problem, we focus on the latter category as these methods share the goal of ultimately producing a sharp deblurred result.

A common strategy for defocus deblurring is to first compute a defocus map and use that information to guide the deblurring.  Defocus map estimation methods~\cite{d2016non,karaali2017edge,lee2019deep,park2017unified,shi2015just} estimate the amount of defocus blur per pixel for an image with defocus blur.   Representative works include Karaali et al.~\cite{karaali2017edge}, which uses image gradients to calculate the blur amount difference between the original image edges and their re-blurred ones. Park et al.~\cite{park2017unified} introduced a method based on hand-crafted and deep features that were extracted from a pre-trained blur classification network. The combined feature vector was fed to a regression network to estimate the blur amount on edges and then later deblur the image. Shi et al.~\cite{shi2015just} proposed an effective blur feature using a sparse representation and image decomposition to detect just noticeable blur.   Methods that directly deblur the image include Andr{\`e}s et al.'s~\cite{d2016non} approach, which uses regression trees to deblur the image.  Recent work by Lee et al.~\cite{lee2019deep} introduced a DNN architecture to estimate an image defocus map using a domain adaptation approach. This approach also introduced the first large-scale dataset for DNN-based training.  Our work is inspired by Lee et al.'s~\cite{lee2019deep} success in applying DNNs for the DoF deblurring task.  Our distinction from the prior work is the use of the DP sensor information available at capture time.

\noindent{\textbf{Defocus blur datasets.}}~There are several datasets available for defocus deblurring.  The CUHK~\cite{shi2014discriminative} and DUT~\cite{zhao2018defocus} datasets have been used for blur detection and provide real images with their corresponding binary masks of blur/sharp regions. The SYNDOF~\cite{lee2019deep} dataset provided data for defocus map estimation, in which their defocus blur is synthesized based on a given depth map of pinhole image datasets. The datasets of~\cite{lee2019deep,shi2014discriminative,zhao2018defocus} do not provide the corresponding ground truth all-in-focus image. The RTF~\cite{d2016non} dataset provided  light-field images captured by a Lytro camera for the task of defocus deblurring. In their data, each blurred image has a corresponding all-in-focus image. However, the RTF dataset is small, with only 22 image pairs. While there are other similar and much larger light-field datasets~\cite{hazirbas2018deep,srinivasan2017learning}, these datasets were introduced for different tasks (i.e., depth from focus and synthesizing a 4D RGBD light field), which are different from the task of this paper. In general, the images captured by Lytro cameras are not representative of DSLR and smartphone cameras, because they apply synthetic defocus blur, and have a relatively small spatial resolution~\cite{boominathan2014improving}.

As our approach is to utilize the DP data for defocus deblurring, we found it necessary to capture a new dataset. Our DP defocus blur dataset provides 500 pairs of images of unrepeated scenes; each pair has a defocus blurred image with its corresponding sharp image. The two DP views of the blurred image are also provided, resulting in a total of 2000 images.  Details of our dataset capture are provided in Sec.~\ref{sec:dataCollection}.  Similar to the patch-wise training approach followed in~\cite{lee2019deep,park2017unified}, we extract a large number of image patches from our dataset to train our DNN.

\noindent{\textbf{DP sensor applications.}}~The DP sensor design was developed by Canon for the purpose of optimizing camera autofocus. DP sensors perform what is termed {\it phase difference autofocus} (PDAF)~\cite{abuolaim2018revisiting,abuolaim2020online,jang2015sensor}, in which the phase difference between the left and right sub-aperture views of the primary lens is calculated to measure the blur amount.  Using this phase information, the camera's lens is adjusted such that the blur is minimized. While intended for autofocus, the DP images have been found useful for other tasks, such as depth map estimation~\cite{punnappurath2020modeling,garg2019learning}, reflection removal~\cite{punnappurath2019reflection}, and synthetic DoF~\cite{wadhwa2018synthetic}. Our work is inspired by these prior methods and examines the use of DP data for the task of defocus blur removal.

\begin{figure}[t]
\includegraphics[width=\linewidth]{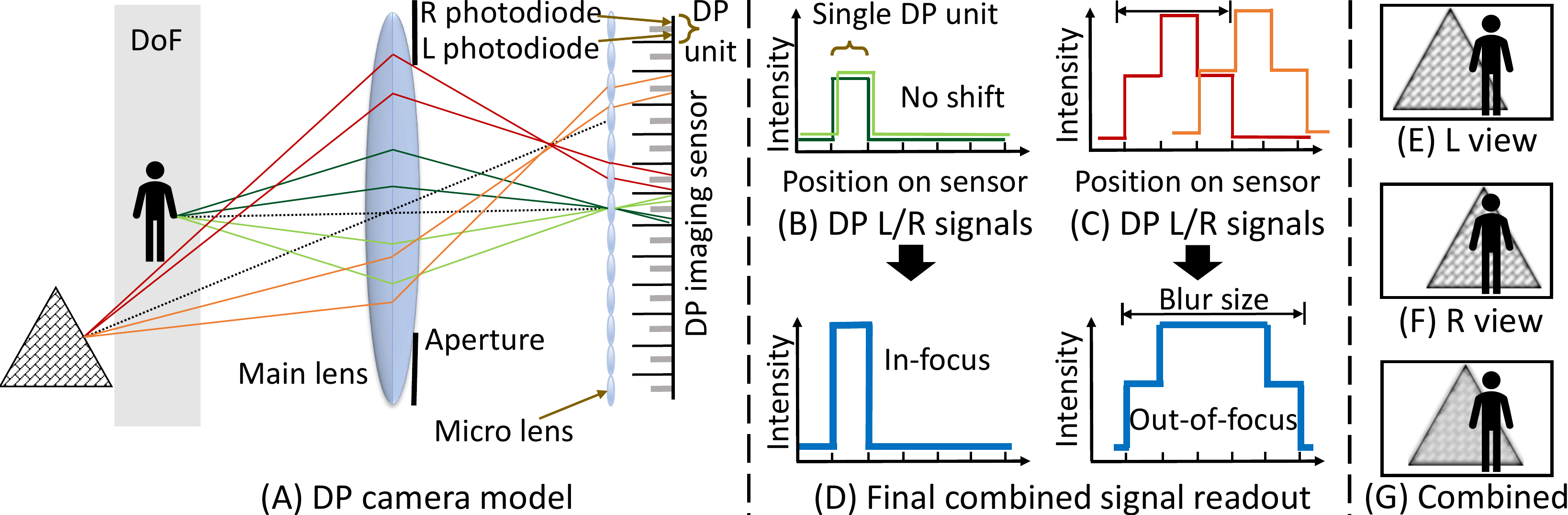}
\caption{Image formation diagram for a DP sensor. (A) Shows a thin-lens camera and a DP sensor. The light rays from different halves of the main lens fall on different left and right photodiodes. (B) Scene points that are within the DoF (highlighted in gray) have no phase shift between their L/R views.  Scene points outside DoF have a phase shift as shown in (C). The L/R signals are aggregated and the corresponding combined signal is shown in (D). The blur size of the L signal is smaller than the combined one in the out-of-focus case. The defocus disparity is noticeable between the captured L/R images (see (E) and (F)). The final combined image in (G) has more blur.  Our DNN leverages this additional information available in the L/R views for image defocus deblurring.}\label{fig:DP}
\end{figure}

\section{DP image formation}

\begin{figure}[t]
\includegraphics[width=\linewidth]{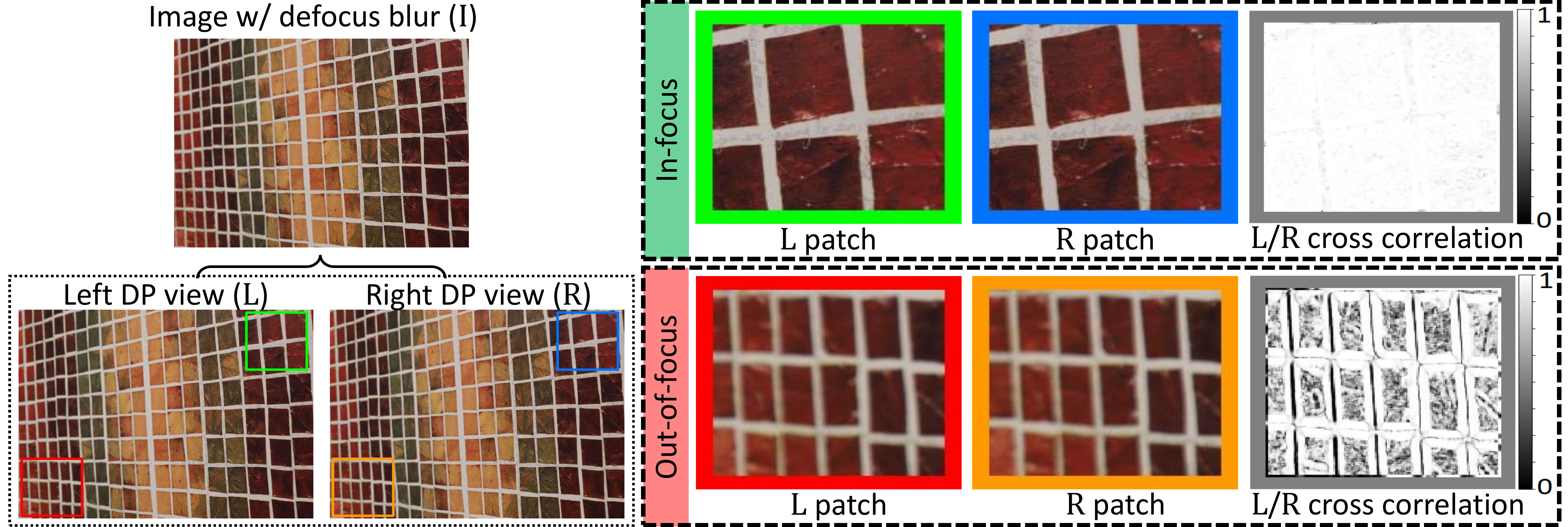}
\caption{An input image $\mathrm{I}$ is shown with a spatially varying defocus blur.  The two dual-pixel (DP) images ($\mathrm{L}$ and $\mathrm{R}$) corresponding to $\mathrm{I}$ are captured at imaging time. In-focus and out-of-focus patches in the  $\mathrm{L}$ and $\mathrm{R}$ DP image patches exhibit different amounts of pixel disparity as shown by the cross-correlation of the two patches.  This information helps the DNN to learn the extent of blur in different regions of the image.}\label{fig:information}
\end{figure}

We begin with a brief overview of the DP image formation. As previously mentioned, the DP sensor was designed to improve camera auto-focus technology.  Fig.~\ref{fig:DP} shows an illustrative example of how DP imaging works and how the left/right images are formed. A DP sensor provides a pair of photodiodes for each pixel with a microlens placed at the pixel site, as shown in Fig.~\ref{fig:DP}-A. This DP unit arrangement allows each pair of photodiodes (i.e., dual-pixel) to record the light rays independently. Depending on the sensor’s orientation, this arrangement can be shown as left/right or top/down pair; in this paper, we refer to them as the left/right pair -- or L and R.  The difference between the two views is related to the defocus amount at that scene point, where out-of-focus scene points will have a difference in phase and be blurred in opposite directions using a point spread function (PSF) and its  flipped one~\cite{punnappurath2020modeling}. This difference yields noticeable defocus disparity that is correlated to the amount of defocus blur.

The phase-shift process is illustrated in Fig.~\ref{fig:DP}.  The person shown in Fig.~\ref{fig:DP}-A is within the camera's DoF, as highlighted in gray, whereas the textured pyramid is outside the DoF. The light rays from the in-focus object converge at a single DP unit on the imaging sensor, resulting in an in-focus pixel and no disparity between their DP L/R views (Fig.~\ref{fig:DP}-B). The light rays coming from the out-of-focus regions spread across multiple DP units and therefore produce a difference between their DP L/R views, as shown in Fig.~\ref{fig:DP}-C.  Intuitively, this information can be exploited by a DNN to learn where regions of the image exhibit blur and the extent of this blur.  The final output image is a combination of the L/R views, as shown in Fig.~\ref{fig:DP}-G.

By examining real examples shown in Fig.~\ref{fig:information} it becomes apparent how a DNN can leverage these two sub-aperture views as input to deblur the image.  In particular, patches containing regions that are out-of-focus will exhibit a notable defocus disparity in the two views that is directly correlated to the amount of defocus blur.  By training a DNN with sufficient examples of the L/R views and the corresponding all-in-focus image, the DNN can learn how to detect and correct blurred regions. Animated examples of the difference between the DP views are provided in the supplemental materials.

\section{Dataset collection}~\label{sec:dataCollection}

Our first task is to collect a dataset with the necessary DP information for training our DNN.  While most consumer cameras employ PDAF sensors, we are aware of only two camera manufacturers that provide DP data -- Google and Canon.  Specifically, Google's research team has released an application to read DP data~\cite{google2019api} from the Google Pixel 3 and 4 smartphones. However, smartphone cameras are currently not suitable for our problem for two reasons. First, smartphone cameras use fixed apertures that cannot be adjusted for data collection.  Second, smartphone cameras have narrow aperture and exhibit large DoF; in fact, most cameras go to great lengths to simulate shallow DoF by purposely introducing defocus blur~\cite{wadhwa2018synthetic}. As a result, our dataset is captured using a Canon EOS 5D Mark IV DSLR camera, which provides the ability to save and extract full-frame DP images.

\begin{figure}[t]
\includegraphics[width=\linewidth]{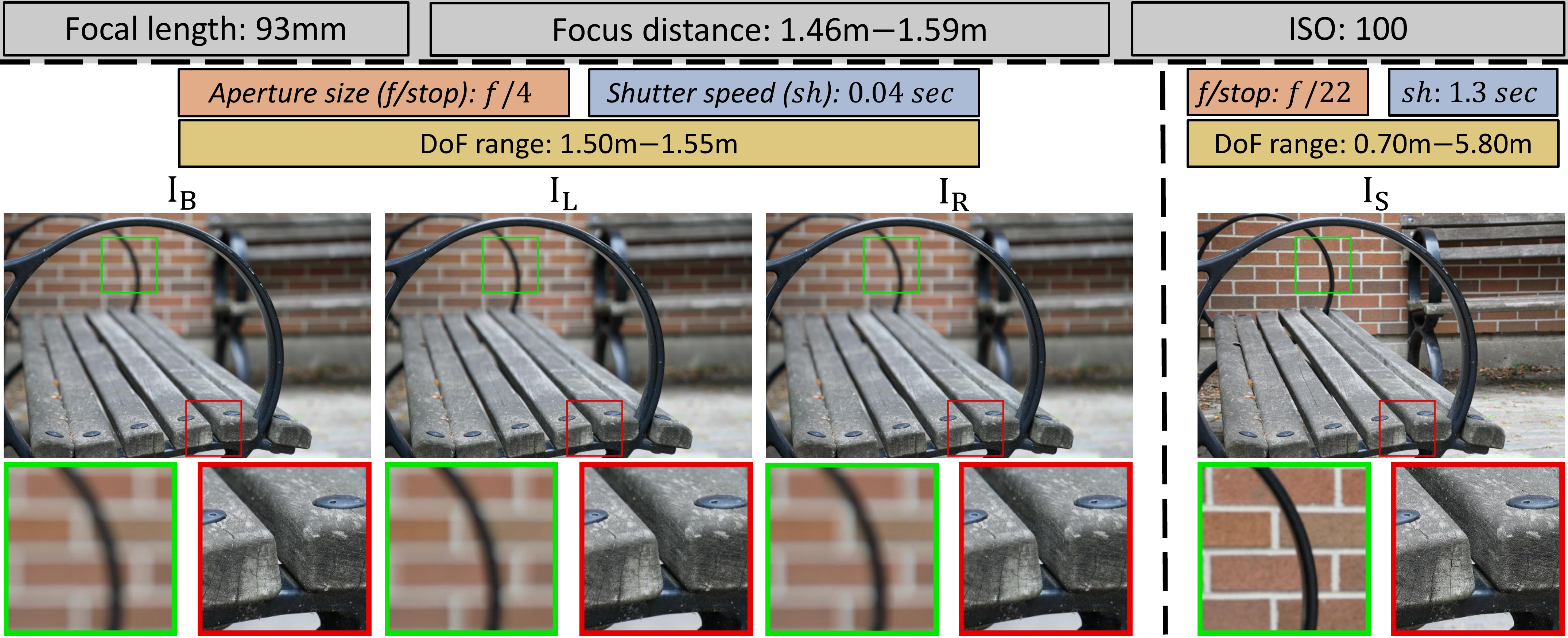}
\caption{An example of an image pair with the camera settings used for capturing. $\mathrm{I_L}$ and $\mathrm{I_R}$ represent the Left and Right DP views extracted from $\mathrm{I_B}$. The focal length, ISO, and focus distance are fixed between the two captures of $\mathrm{I_B}$ and $\mathrm{I_S}$. The aperture size is different, and hence the shutter speed and DoF are accordingly different too. In-focus and out-of-focus zoomed-in patches are extracted from each image and shown in green and red boxes, respectively.}\label{fig:capturingSettings}
\end{figure}

Using the Canon camera, we capture a pair of images of the same static scene at two aperture sizes -- $f/4$ and $f/22$ -- which are the maximum (widest) and minimum (narrowest) apertures possible for our lens configuration. The lens position and focal length remain fixed during image capture. Scenes are captured in aperture-priority mode, in which the exposure compensation between the  image pairs is done automatically by adjusting the shutter speed. The image captured at $f/4$ has the smallest DoF and results in the blurred input image $\mathrm{I_B}$. The image captured at $f/22$ has the largest DoF and serves as the all-in-focus target image denoted as $\mathrm{I_S}$ (sharp image). Focus distance and focal length differ across captured pairs in order to capture a diverse range of defocus blur types. Our captured images offer the following benefits over prior datasets:\\
\noindent{\textbf{High-quality images.}}~Our captured images are low-noise images (i.e., low ISO equates to low-noise~\cite{plotz2017benchmarking}) and of full resolution of $6720\times4480$.  All images, including the left/right DP views, are processed to an sRGB and encoded with a lossless 16-bit depth per RGB channel.\\
\noindent{\textbf{Real and diverse defocus blur.}}~Unlike other existing datasets, our dataset provides real defocus blur and in-focus pairs indicative of real camera optics.\\
\noindent{\textbf{Varying scene contents.}}~To provide a wide range of object categories, we collect 500 pairs of unique indoor/outdoor scenes with a large variety of scene contents. Our dataset is also free of faces to avoid privacy issues.

The $f/4$ (blurry) and $f/22$ (sharp) image pairs are carefully imaged static scenes with the camera fixed on a tripod. To further avoid camera shake, the camera was controlled remotely to allow hands-free operation.  Fig.~\ref{fig:capturingSettings} shows an example of an image pair from our dataset. The left and right DP views of $\mathrm{I_B}$ are provided by the camera and denoted as $\mathrm{I_L}$ and $\mathrm{I_R}$ respectively. The ISO setting is fixed for each image pair. Fig.~\ref{fig:capturingSettings} shows the DP L/R views for only image $\mathrm{I_B}$, because DP L/R views of $\mathrm{I_S}$ are visually identical due to the fact $\mathrm{I_S}$ is our all-in-focus ground truth.

\section{Dual-pixel defocus deblurring DNN (DPDNet)}~\label{sec:methodology}

Using our captured dataset, we trained a symmetric encoder-decoder CNN architecture with skip connections between the corresponding feature maps~\cite{mao2016image,ronneberger2015u}. Skip connections are widely used in encoder-decoder CNNs to combine various levels of feature maps. These have been found useful for gradient propagation and convergence acceleration and to allow training of deeper networks as stated in~\cite{he2016deep,srivastava2015training}.

We adapt a U-Net-like architecture~\cite{ronneberger2015u} with the following modifications:  an input layer to take a 6-channel input cube (two DP views; each is a 3-channel sRGB image) and an output layer to generate a 3-channel output sRGB image; skip connections of the convolutional feature maps are passed to their mirrored convolutional layers without cropping in order to pass on more feature map detail; and the loss function is changed to be mean squared error (MSE).

\begin{figure}[t]
\includegraphics[width=\linewidth]{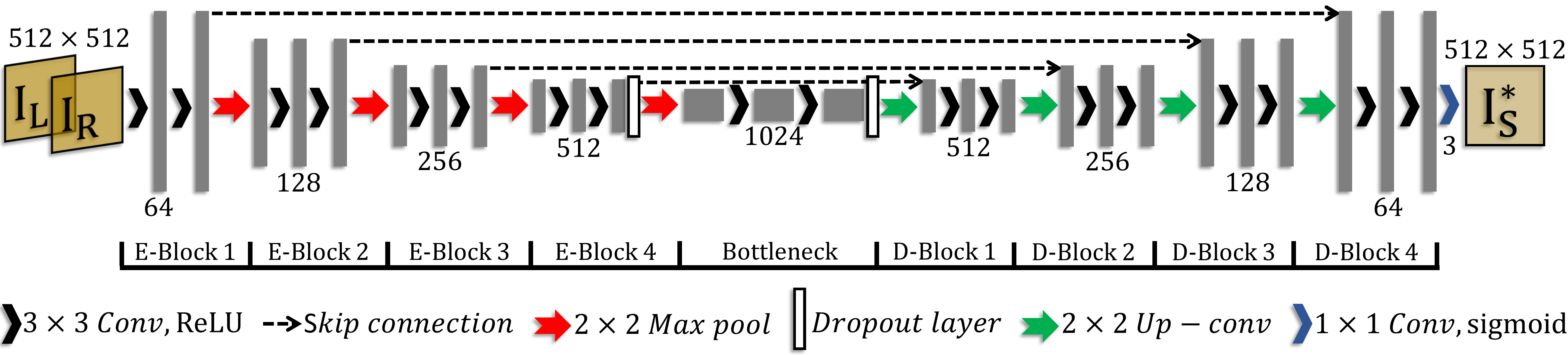}
\caption{Our proposed DP deblurring architecture (DPDNet). Our method utilizes the DP images, $\mathrm{I_L}$ and $\mathrm{I_R}$, for predicting the sharp image $\mathrm{I_S^*}$ through three stages: encoder (E-Blocks), bottleneck, and decoder (D-Blocks). The size of the input and output layers is shown above the images. The number of output filters is shown under the convolution operations for each block.}\label{fig:DBDframework}
\end{figure}

The overall DNN architecture of our proposed DP deblurring method is shown in Fig.~\ref{fig:DBDframework}. Our method reads the two DP images, $\mathrm{I_L}$ and $\mathrm{I_R}$, as a 6-channel cube, and processes them through the encoder, bottleneck, and decoder stages to get the final sharp image $\mathrm{I_S^*}$. There are four blocks in the encoder stage (E-Block 1--4) and in each block, two $3\times3$ convolutional operations are performed, each followed by a ReLU activation. Then a $2\times2$ max pooling is performed for downsampling. Although max pooling operations reduce the size of feature maps between E-Blocks, this is required to extend the receptive field size in order to handle large defocus blur. To reduce the chances of overfitting, two dropout layers are added, one before the max pooling operation in the fourth E-Block, and one dropout layer at the end of the network bottleneck, as shown in Fig.~\ref{fig:DBDframework}. In the decoder stage, we also have four blocks (D-Block 1--4). For each D-Block, a $2\times2$ upsampling of the input feature map followed by a $2\times2$ convolution ($Up-conv$) is carried out instead of direct deconvolution in order to avoid checkerboard artifacts~\cite{odena2016deconvolution}. The corresponding feature map from the encoder stage is concatenated. Next, two $3\times3$ convolutions are performed, each followed by a ReLU activation. Afterwards, a $1\times1$ convolution followed by sigmoid activation is applied to output the final sharp image $\mathrm{I_S^*}$. The number of output filters is shown under each convolution layer for each block in Fig.~\ref{fig:DBDframework}. The stride for all operations is $1$ except for the max pooling operation, which has a stride of $2$. The final sharp image $\mathrm{I_S^*}$ is, thus, predicted as follows:
\begin{equation}
\mathrm{I_S^*}=\textrm{DPDNet}(\mathrm{I_L},\mathrm{I_R};\theta_\textrm{DPDNet}),
\end{equation}
where DPDNet is our proposed architecture, and $\theta_\textrm{DPDNet}$ is the set of weights and parameters.

\noindent{\textbf{Training procedure.}}~
The size of input and output layers is set to $512\times512\times6$ and $512\times512\times3$, respectively.  This is because we train not on the full-size images but on the extracted image patches. We adopt the weight initialization strategy proposed by He~\cite{he2015delving} and use the Adam optimizer~\cite{kingma2014adam} to train the model. The initial learning rate is set to $2\times10^{-5}$, which is decreased by half every 60 epochs. We train our model with mini-batches of size 5 using MSE loss between the output and the ground truth as follows:
\begin{equation}
\mathcal{L}=\frac{1}{n}\sum_n(\mathrm{I_S}-\mathrm{I_S^*})^2,
\end{equation}
where $n$ is the size of the image patch in pixels. During the training phase, we set the dropout rate to $0.4$. All the models described in the subsequent sections are implemented using Python with the Keras framework on top of TensorFlow and trained with a NVIDIA TITAN X GPU. We set the maximum number of training epochs to 200.

\section{Experimental results}\label{sec:results}

\begin{table}[t]
\centering
\newcommand{\cl}{35}
\resizebox{\linewidth}{!}
{
\begin{tabular}{c | c|c|c|c || c|c|c|c || c|c|c|c}
\toprule
\multirow{2}{*}{\bf Method} &\multicolumn{4}{c}{\bf Indoor} &\multicolumn{4}{c}{\bf Outdoor} &\multicolumn{4}{c}{\bf Combined}\\ \cline{2-13}
				
		& PSNR $\uparrow$ & SSIM $\uparrow$ & MAE $\downarrow$ & LPIPS $\downarrow$ & PSNR $\uparrow$ & SSIM $\uparrow$ & MAE $\downarrow$ & LPIPS $\downarrow$ & PSNR $\uparrow$ & SSIM $\uparrow$ & MAE $\downarrow$ & LPIPS $\downarrow$ \\ \midrule \midrule
				
{\bf EBDB~\cite{karaali2017edge}}	& 25.77 & 0.772 &0.040 & 0.297 & 21.25 & 0.599 & 0.058 & 0.373 & 23.45 & 0.683 & 0.049 & 0.336 \\ \midrule
{\bf DMENet~\cite{lee2019deep}}		& 25.50 & 0.788 & 0.038 & 0.298 & 21.43 & 0.644 & 0.063 & 0.397 & 23.41 & 0.714 & 0.051 & 0.349 \\ \midrule
{\bf JNB~\cite{shi2015just}} 	& \cellcolor{blue!15}26.73 & \cellcolor{blue!15}0.828 & \cellcolor{blue!15}0.031 &0.273 & 21.10 & 0.608 & 0.064 & 0.355 & 23.84 & 0.715 & 0.048 & 0.315 \\ \midrule
{\bf Our $\textrm{DPDNet}$-Single}	 	& 26.54 & 0.816 & \cellcolor{blue!15}0.031 & \cellcolor{blue!15}0.239 & \cellcolor{blue!15} 22.25 & \cellcolor{blue!15} 0.682 & \cellcolor{blue!15} 0.056 & \cellcolor{blue!15} 0.313 & \cellcolor{blue!15}24.34 & \cellcolor{blue!15}0.747 & \cellcolor{blue!15}0.044 & \cellcolor{blue!15}0.277 \\ \midrule
{\bf Our DPDNet}	 	& \cellcolor{green!25}{\bf 27.48} & \cellcolor{green!25}{\bf 0.849} & \cellcolor{green!25}{\bf 0.029} & \cellcolor{green!25}{\bf 0.189} & \cellcolor{green!25}{\bf 22.90} & \cellcolor{green!25}{\bf 0.726} & \cellcolor{green!25}{\bf 0.052} & \cellcolor{green!25}{\bf 0.255} & \cellcolor{green!25}{\bf 25.13} & \cellcolor{green!25}{\bf 0.786} & \cellcolor{green!25}{\bf 0.041} & \cellcolor{green!25}{\bf 0.223} \\ \midrule
\end{tabular}
}
\caption{The quantitative results for different defocus deblurring methods. The testing on the dataset is divided into three scene categories: indoor, outdoor, and combined. The top result numbers are highlighted in green and the second top in blue.
DPDNet-Single is our DPDNet variation that is trained with only a single blurred input. Our DPDNet that uses the two L/R DP views achieved the best results on all scene categories for all metrics. Note: the testing set consists of 37 indoor and 39 outdoor scenes.}
\label{tab:quantitative}
\end{table}

We first describe our data preparation procedure and then evaluation metrics used. This is followed by quantitative and qualitative results to evaluate our proposed method with existing deblurring methods. We also discuss the time analysis and test the robustness of our DP method against different aperture settings.

\noindent{\textbf{Data preparation.}}~Our dataset has an equal number of indoor and outdoor scenes. We divide the data into $70\%$ training, $15\%$ validation, and $15\%$ testing sets. Each set has a balanced number of indoor/outdoor scenes. To prepare the data for training, we first downscale our images to be $1680\times1120$ in size. Next, image patches are extracted by sliding a window of size $512\times512$ with $60\%$ overlap.   We empirically found this image size and patch size to work well. An ablation study of different architecture settings is provided in  the supplemental materials.
We compute the sharpness energy (i.e., by applying Sobel filter) of the in-focus image patches and sort them. We discard 30\% of the patches that have the lowest sharpness energy.  Such patches represent homogeneous regions, cause an ambiguity associated to the amount of blur, and adversely affect the DNNs training, as found in~\cite{park2017unified}.

\begin{figure}[p]
\includegraphics[width=0.97\linewidth]{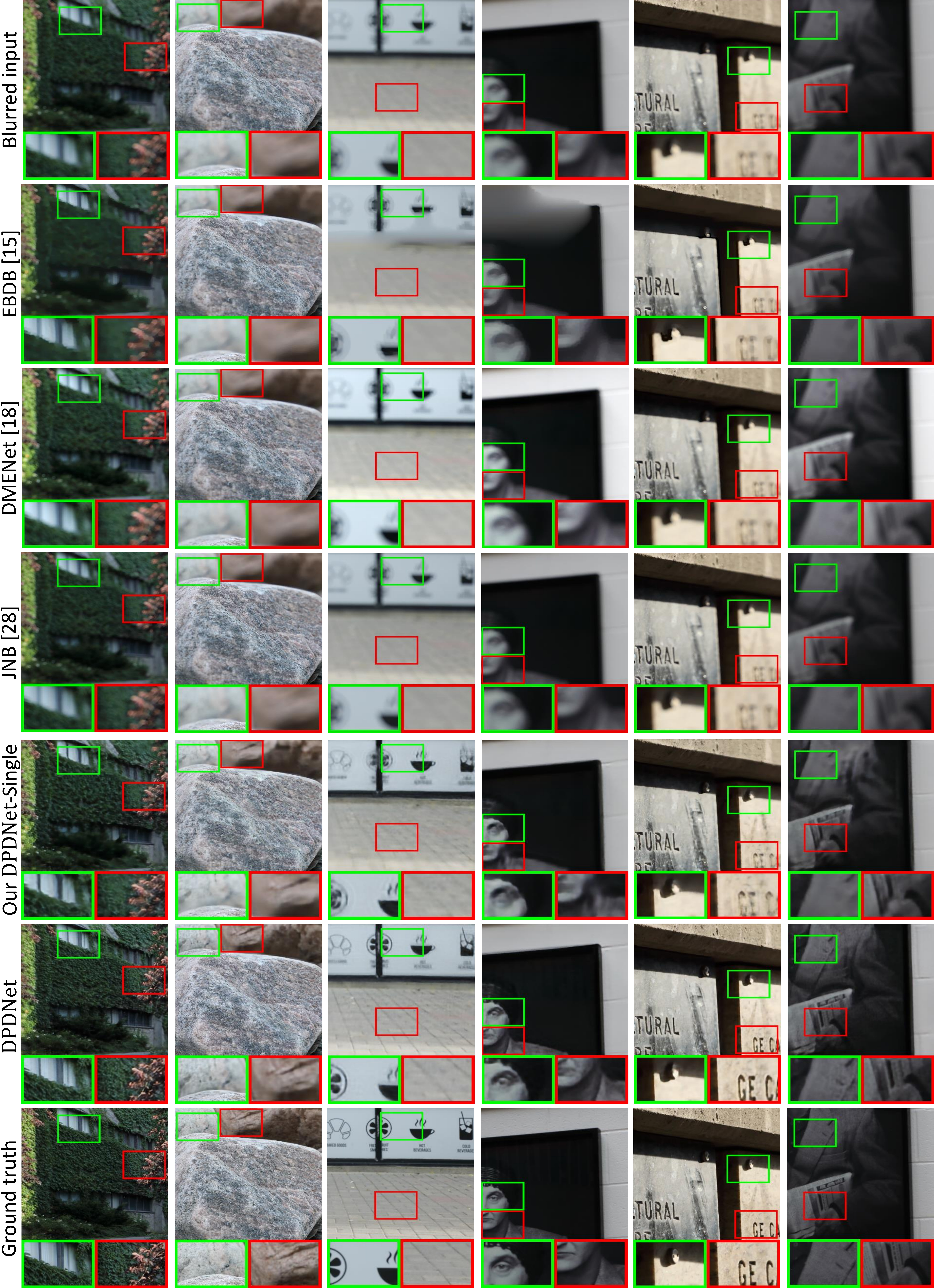}
\caption{Qualitative comparisons of different deblurring methods. The first row is the input image that has a spatially varying blur, and the last row is the corresponding ground truth sharp image. The rows in between are the results of different methods. We also present zoomed-in cropped patches in green and red boxes. Our DPDNet method significantly outperforms other methods in terms of deblurring quality.}\label{fig:qaulRes}
\end{figure}

\begin{figure}[t]
\centering
\includegraphics[width=\linewidth]{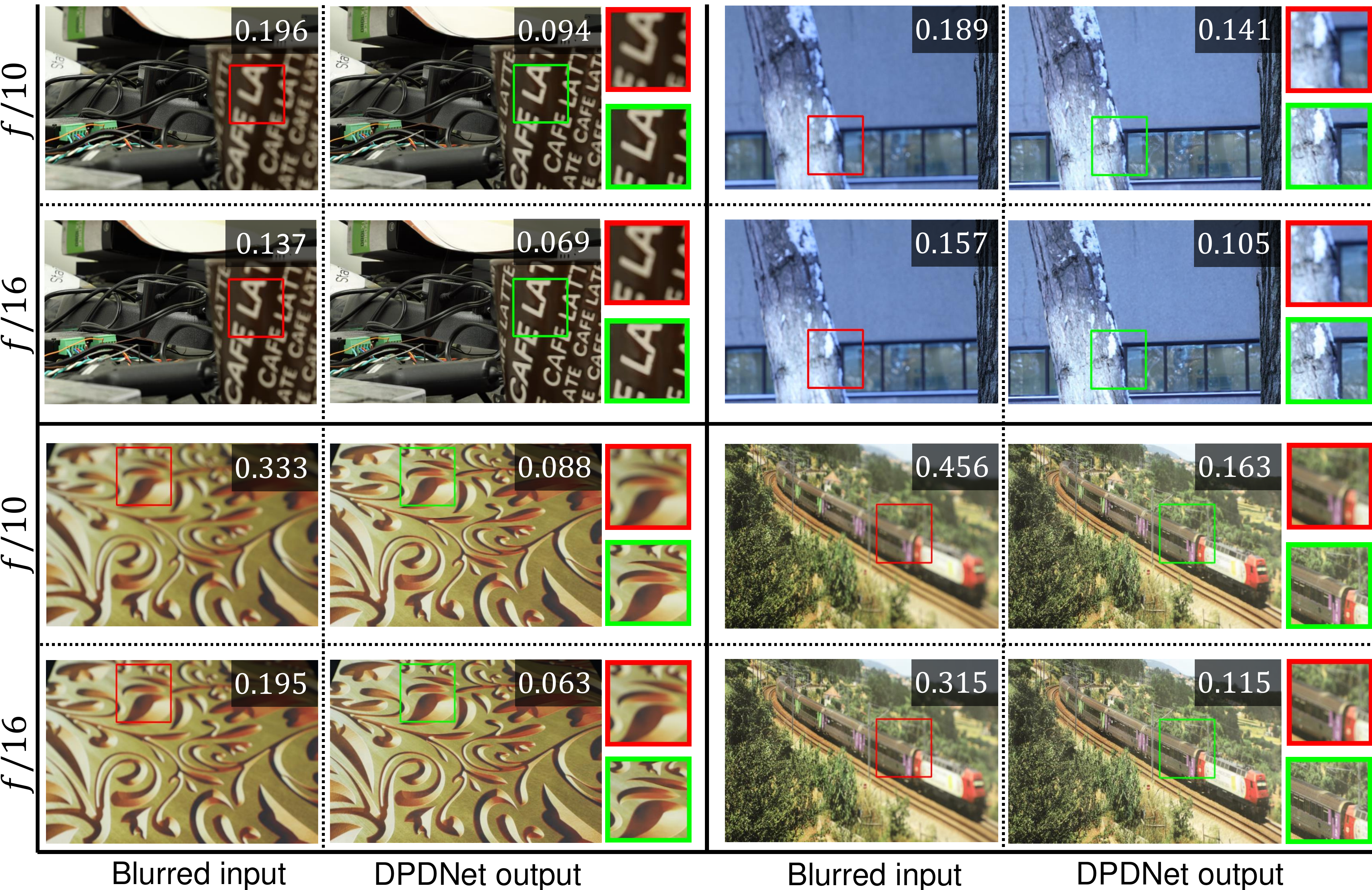}
\caption{Examining DPDNet's robustness to different aperture settings. Four scenes are presented; each has two different apertures. In each scene, the left-hand image is the blurred one, $\mathrm{I_B}$, and the right-hand image is the deblurred one, $\mathrm{I_S^*}$, computed by our DPDNet. The number shown on each image is the LPIPS measure compared with the ground truth $\mathrm{I_S}$. Zoomed-in cropped patches are also provided. Even though our training data was on blurry examples with an $f/4$ aperture, our DPDNet is able to generalize well to different apertures.}\label{fig:DDBDRobust}
\end{figure}

\noindent{\textbf{Evaluation metrics.}}~Results are reported on traditional signal processing metrics -- namely, PSNR, SSIM~\cite{wang2004image}, and MAE.  We also incorporate the recent learned perceptual image patch similarity  (LPIPS) proposed by~\cite{zhang2018unreasonable}. The LPIPS metric is correlated with human perceptual similarity judgments as a perceptual metric for low-level vision tasks, such as enhancement and image deblurring.

\noindent{\textbf{Quantitative results.}}~We compare our DPDNet with the following three methods: the edge-based defocus blur (EBDB)~\cite{karaali2017edge}, the defocus map estimation network (DMENet)~\cite{lee2019deep}, and the just noticeable blur (JNB)~\cite{shi2015just} estimation.  These methods accept only a single image as input -- namely, $\mathrm{I_B}$ -- and estimate the defocus map in order to use it to guide the deblurring process. The EBDB~\cite{karaali2017edge} and JNB~\cite{shi2015just} are not learning-based methods. We test them directly on our dataset using $\mathrm{I_B}$ as input. The EBDB uses a combination of non-blind deblurring methods proposed in~\cite{krishnan2009fast,levin2007image}, and for a fair comparison, we contacted the authors for their deblurring settings and implementation. The JNB method uses the non-blind defocus deblurring method from~\cite{fish1995blind}.

For the deep-learning-based method (i.e., DMENet~\cite{lee2019deep}), the method requires the ground truth defocus map for training. In our dataset, we do not have this ground truth defocus map and provide only the sharp image, since our approach in this work is to solve directly for defocus deblurring. Therefore, we tested the DMENet on our dataset using $\mathrm{I_B}$ as input without retraining. For deblurring, DMENet adopts a non-blind deconvolution algorithm proposed by~\cite{krishnan2009fast}.  Our results are compared against code provided by the authors.  Unfortunately, the methods in~\cite{d2016non,park2017unified} do not have the deblurring code available for comparison.

To show the advantage of utilizing DP data for defocus deblurring, we introduce a variation of our DPDNet that accepts only a single input (i.e., $\mathrm{I_B}$) and uses exactly the same architecture settings along with the same training procedure as shown in Fig~\ref{fig:DBDframework}. We refer to this variation as $\textrm{DPDNet}$-Single in Table~\ref{tab:quantitative}.  Our proposed architecture is fully convolutional, which enables testing any image size during the testing phase. Therefore, all the subsequent results are reported on the testing set using the full image for all methods.
Table~\ref{tab:quantitative} reports our findings by testing on three scene categories: indoor, outdoor, and combined.  Top result numbers are highlighted in green and the second top ones in blue. Our DPDNet method has a significantly better deblurring ability based on all metrics for all testing categories. Furthermore, DP data is the key that made our DPDNet method outperforms others, especially the single image input one (i.e., $\textrm{DPDNet}$-Single), in which it has exactly the same architecture but does not utilize DP views. Interestingly, all methods have better deblurring results for indoor scenes, due to the fact that outdoor scenes tend to have larger depth variations, and thereby more defocus blur.

\noindent{\textbf{Qualitative results.}}~In Fig.~\ref{fig:qaulRes}, we present the qualitative results of different defocus deblurring methods. The first row shows the input image with a spatially varying defocus blur; the last row shows the corresponding ground truth sharp image. The rows in between present different methods, including ours. This figure also shows two zoomed-in cropped patches in green and red to further illustrate the difference visually. From the visual comparison with other methods, our DPDNet has the best deblurring ability and is quite similar to the ground truth. EBDB~\cite{karaali2017edge}, DMENet~\cite{lee2019deep}, and JNB~\cite{shi2015just} are not able to handle spatially varying blur with almost unnoticeable difference with the input image. EBDB~\cite{karaali2017edge} tends to introduce some artifacts in some cases. Our single image method (i.e., $\textrm{DPDNet}$-Single) has better deblurring ability compared to other traditional deblurring methods, but it is not at the level of our method that utilizes DP views for deblurring. Our DPDNet method, as shown visually, is effective in handling spatially varying blur. For example, in the second row, the image has a part that is in focus and another is not; our DPDNet method is able to determine the deblurring amount required for each pixel, in which the in-focus part is left untouched. Further qualitative results are provided in our supplemental materials, including results on DP data obtained from a smartphone camera.

\noindent{\textbf{Time analysis.}}~We examine evaluating different defocus deblurring methods based on the time required to process a testing image of size $1680\times1120$ pixels. Our DPDNet directly computes the sharp image in a single pass, whereas other methods~\cite{karaali2017edge,lee2019deep,shi2015just} use two passes: (1) defocus map estimation and (2) non-blind deblurring based on the estimated defocus map.

Non-learning-based methods (i.e., EBDB~\cite{karaali2017edge} and JNB~\cite{shi2015just}) do not utilize the GPU and use only the CPU. For the deep-learning method (i.e., DMENet~\cite{lee2019deep}), it utilizes the GPU for the first pass; however, the deblurring routine is applied on a CPU. This time evaluation is performed using Intel Core i7-6700 CPU and NVIDIA TITAN X GPU. Our DPDNet operates in a single pass and can process the testing image of size $1680\times1120$ pixels about $1.2\times10^3$ times faster compared to the second-best method (i.e., DMENet), as shown in Table~\ref{tab:time}.

\begin{table*}[t]
\centering
\newcommand{\cl}{35}
\resizebox{0.6\linewidth}{!}
{
\begin{tabular}{c | c|c|c}
\toprule
\multirow{2}{*}{\bf Method} &\multicolumn{3}{c}{\bf Time (Sec) $\downarrow$} \\ \cline{2-4}
				
		& Defocus map estimation & Defocus deblurring & Total \\ \midrule \midrule
{\bf EBDB~\cite{karaali2017edge}}	&  57.2 &  872.5 & 929.7\\ \midrule
{\bf DMENet~\cite{lee2019deep}}		&  1.3 &  612.4 & 613.7\\ \midrule
{\bf JNB~\cite{shi2015just}}		&  605.4 &  237.7 & 843.1\\ \midrule
{\bf Our DPDNet}	 					& {\bf 0} &  {\bf 0.5} & {\bf 0.5}\\ \midrule
\end{tabular}
}
\caption{Time analysis of different defocus deblurring methods. The last column is the total time required to process a testing image of size $1680\times1120$ pixels. Our DPDNet is about $1.2\times10^3$ times faster compared to the second-best method (i.e., DMENet).}
\label{tab:time}
\end{table*}

\noindent{\textbf{Robustness to different aperture settings.}}~In our dataset, the image pairs are captured using aperture settings corresponding to f-stops $f/22$ and $f/4$. Recall that $f/4$ results in the greatest DoF and thus most defocus blur. Our  DPDNet is trained on diverse images with many different depth values; thus, our training data spans the worst-case blur that would be observed with any aperture settings. To test the ability of our DPDNet in generalizing for scenes with different aperture settings, we capture image pairs with aperture settings $f/10$ and $f/16$ for the blurred image and again $f/22$ for the corresponding ground truth image. Our DPDNet is applied to these less blurred images.  Fig.~\ref{fig:DDBDRobust} shows the results for four scenes, where each scene's image has its LPIPS measure compared with the ground truth. For better visual comparison, Fig.~\ref{fig:DDBDRobust} provides zoomed-in patches that are cropped from the blurred input (red box) and the deblurred one (green box). These results show that our DPDNet is able to deblur scenes with different aperture settings that have not been used during training.
\section{Applications}
Image blur can have a negative impact on some computer vision tasks, as found in~\cite{guo2019effects}. Here we investigate defocus blur effect on two common computer vision tasks -- namely, image segmentation  and monocular depth estimation.

\begin{figure}[t]
\centering
\includegraphics[width=0.96\linewidth]{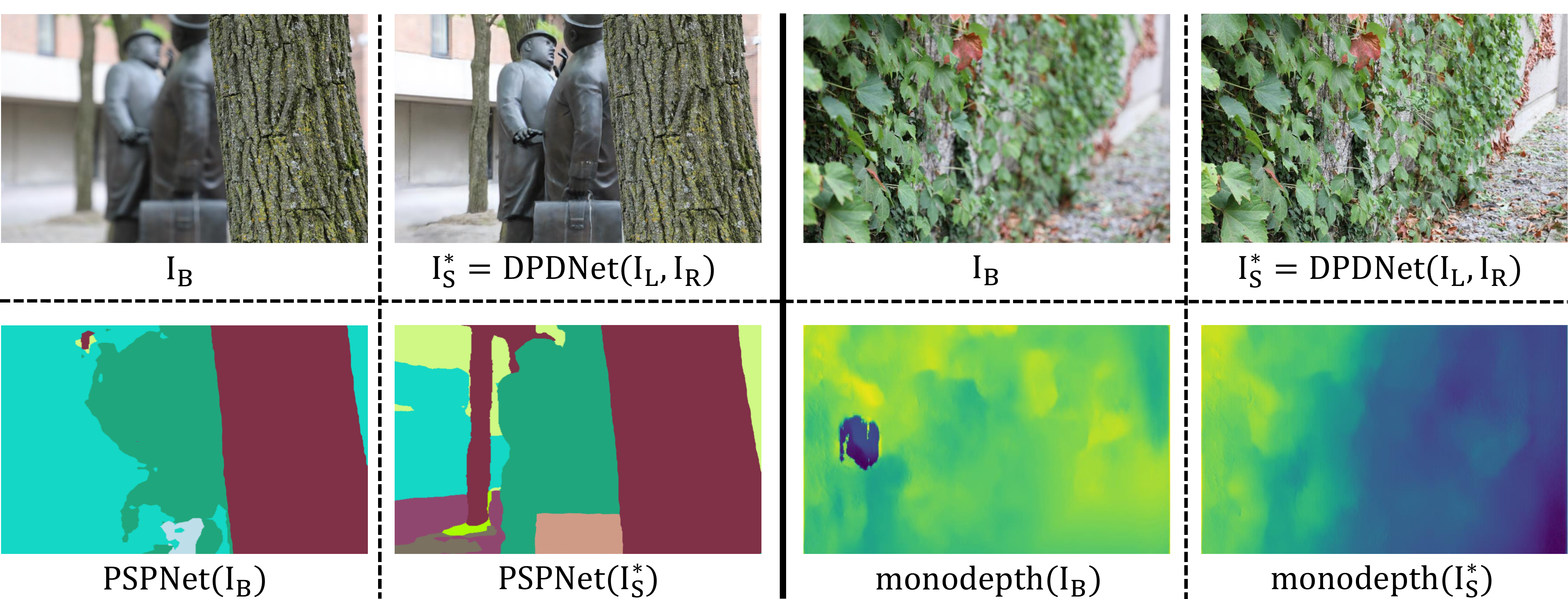}
\caption{The effect of defocus blur on some computer vision tasks. The first two columns show the image segmentation results using the PSPNet~\cite{zhao2017pyramid} segmentation model. The segmentation results are affected by the blurred image $\mathrm{I_B}$, where a large portion is segmented as unknown in cyan. The last two columns show the results of the monocular depth estimation using the monodepth model from~\cite{monodepth17}. The depth estimation is highly affected by the defocus blur and produced wrong results. Deblurring $\mathrm{I_B}$ using our DP deblurring method has significantly improved the results for both tasks.}\label{fig:imgSeg}
\end{figure}
\noindent{\textbf{Image segmentation.}}~The first two columns in Fig.~\ref{fig:imgSeg} demonstrate the negative effect of defocus blur on the task of image segmentation. We use the PSPNet segementation model from~\cite{zhao2017pyramid}, and test two images: one is the blurred input image $\mathrm{I_B}$ and another is the deblurred one $\mathrm{I_S^*}$ using our DPDNet deblurring model. The segmentation results are affected by $\mathrm{I_B}$ -- only the foreground tree was correctly segmented. PSPNet assigns cyan color to unknown categories, where a large portion of $\mathrm{I_B}$ is segmented as unknown. On the other hand, the segmentation results of $\mathrm{I_S^*}$ are much better, in which more categories are segmented correctly. With that said, image DoF deblurring using our DP method can be beneficial for the task of image segmentation.

\noindent{\textbf{Monocular depth estimation.}}~The monocular depth estimation is the task of estimating scene depth using a single image. In the last two columns of Fig.~\ref{fig:imgSeg}, we show the direct effect of defocus blur on this task. We use the monodepth model from~\cite{monodepth17} to test the two images $\mathrm{I_B}$ and $\mathrm{I_S^*}$ in order to examine the change in performance. The result of monodepth is affected by the defocus blur, in which the depth map estimated is completely wrong. Contrarily, the result of monodepth has been significantly improved after testing with the deblurred input image using our DPDNet deblurring model. Therefore, deblurring images using our DPDNet can be useful for the task of monocular depth map estimation.

\section{Conclusion}

We have presented a novel approach to reduce the effect of defocus blur present in images captured with a shallow DoF. Our approach leverages the DP data that is available in most modern camera sensors but currently being ignored for other uses.  We show that the DP images are highly effective in reducing DoF blur when used in a DNN framework.  As part of this effort, we have captured a new image dataset consisting of blurred and sharp image pairs along with their DP images.     Experimental results show that leveraging the DP data provides state-of-the-art quantitative results on both signal processing and perceptual metrics.  We also demonstrate that our deblurring method can be beneficial for other computer vision tasks. We believe our captured dataset and DP-based method are useful for the research community and will help spur additional ideas about both defocus deblurring and applications that can leverage data from DP sensors.
\\
\\
\noindent \textbf{Acknowledgments.}
This study was funded in part by the Canada First Research Excellence Fund for the Vision: Science to Applications (VISTA) programme and an NSERC Discovery Grant. Dr. Brown contributed to this article in his personal capacity as a professor at York University. The views expressed are his own and do not necessarily represent the views of Samsung Research.

\bibliographystyle{splncs04}

\clearpage

\newcommand{\hbAppendixPrefix}{A}
\setcounter{section}{0}
\setcounter{footnote}{0}
\renewcommand{\thesection}{S\arabic{section}}

\begin{centering}
\large{\textbf{Supplemental Materials}}\\~\\~\\
\end{centering}

The supplemental materials provide an ablation study of different variations of our DPDNet in Sec.~\ref{sec:ablationStudy}. Sec.~\ref{sec:discussion} provides a brief discussion about defocus blur and motion blur. Use cases are described in Sec.~\ref{sec:useCases}. Sec.~\ref{sec:pixel} provides results on dual-pixel (DP) data obtained from a smartphone camera.  Sec.~\ref{sec:qualRes} provides additional quantitative results. There are also 14 animated qualitative examples provided in the ``animated\_results'' directory---located at the github project repository\footnote{https://github.com/Abdullah-Abuolaim/defocus-deblurring-dual-pixel}. Furthermore, as mentioned in Sec.~3 of the main paper, we provide animated examples that show the difference between the dual-pixel (DP) views in the ``animated\_dp\_examples'' directory---located at the github project repository\footnotemark[1].
\section{Ablation study}\label{sec:ablationStudy}
In this section, we provide an ablation study of different variations in training our DPDNet with: (1) an extra input image (Sec.~\ref{sec:extraInput}), (2) less E-Blocks and D-Blocks (Sec.~\ref{sec:lessBlock}), (3) different input sizes (Sec.~\ref{sec:inputSize}), (4) different ratios of homogeneous region filtering (Sec.~\ref{sec:filtering}), and (5) different data types (Sec.~\ref{sec:dataType}). This is related to Sec.~5 and  Sec.~6 of the main paper.

\subsection{DPDNet with extra input image}\label{sec:extraInput}

As described in Sec.~5 of the main paper, our DPDNet takes the two dual-pixel L/R views,
$\mathrm{I_L}$ and $\mathrm{I_R}$, as inputs to estimate the sharp image $\mathrm{I_S^*}$. In our dataset, in addition to the L/R views, we also provide the corresponding combined image $\mathrm{I_B}$ that would be outputted by the camera. In this section, we examine training our DPDNet with all three images, namely $\mathrm{I_L}$, $\mathrm{I_R}$, and $\mathrm{I_B}$. We refer to this variation as DPDNet$(\mathrm{I_L},\mathrm{I_R},\mathrm{I_B})$.

Table~\ref{tab:extraInput} shows the results of the three-input DPDNet, DPDNet$(\mathrm{I_L},\mathrm{I_R},\mathrm{I_B})$, vs. the two-input one, DPDNet$(\mathrm{I_L},\mathrm{I_R})$, proposed in the main paper. The results of all metrics are quite similar with a slight difference. Our conclusion is that training and testing the DPDNet with the extra input $\mathrm{I_B}$ provides no noticeable improvement. Such results are expected, since $\mathrm{I_B}$ is a combination of $\mathrm{I_L}$ and $\mathrm{I_R}$.
\begin{table}[H]
\centering
\newcommand{\cl}{35}
\resizebox{\linewidth}{!}
{
\begin{tabular}{c | c|c|c|c || c|c|c|c || c|c|c|c}
\toprule
\multirow{2}{*}{\bf Method} &\multicolumn{4}{c}{\bf Indoor} &\multicolumn{4}{c}{\bf Outdoor} &\multicolumn{4}{c}{\bf Combined}\\ \cline{2-13}
				
		& PSNR $\uparrow$ & SSIM $\uparrow$ & MAE $\downarrow$ & LPIPS $\downarrow$ & PSNR $\uparrow$ & SSIM $\uparrow$ & MAE $\downarrow$ & LPIPS $\downarrow$ & PSNR $\uparrow$ & SSIM $\uparrow$ & MAE $\downarrow$ & LPIPS $\downarrow$ \\ \midrule \midrule
		
{\bf DPDNet$(\mathrm{I_L},\mathrm{I_R},\mathrm{I_B})$}	 	&  27.32 & 0.842 & {\bf 0.029} & 0.191 & {\bf 22.94} & 0.723 & {\bf 0.052} & 0.257 & 25.07 & 0.781 & {\bf 0.041} & 0.225 \\ \midrule
{\bf DPDNet$(\mathrm{I_L},\mathrm{I_R})$}	 	& {\bf 27.48} & {\bf 0.849} & {\bf 0.029} & {\bf 0.189} & 22.90 & {\bf 0.726} & {\bf 0.052} & {\bf 0.255} & {\bf 25.13} & {\bf 0.786} & {\bf 0.041} & {\bf 0.223} \\ \midrule
\end{tabular}
}
\caption{DPDNet with extra input image. The quantitative results of DPDNet$(\mathrm{I_L},\mathrm{I_R},\mathrm{I_B})$ vs. DPDNet$(\mathrm{I_L},\mathrm{I_R},)$ using four metrics. The testing on the dataset is divided into three scene categories: indoor, outdoor, and combined. The best results are in bold numbers. The results of DPDNet$(\mathrm{I_L},\mathrm{I_R},\mathrm{I_B})$ and DPDNet$(\mathrm{I_L},\mathrm{I_R},)$ are quite similar with a slight difference. Note: the testing set consists of 37 indoor and 39 outdoor scenes.}
\label{tab:extraInput}
\end{table}

\subsection{DPDNet with less blocks}\label{sec:lessBlock}

In this section, we train a ``lighter'' version of our DPDNet with less E-Blocks and D-Blocks. This is done by reducing E-Block 1 and D-Block 4. We refer to this light version as DPDNet-Light. In Table~\ref{tab:lessBlock}, we provide a comparison of DPDNet-Light and our full DPDNet that is proposed in the main paper.

Table~\ref{tab:lessBlock} shows that our full DPDNet has a better performance compared to the lighter one. Nevertheless, the sacrifice in performance is not too significant, which implies that the DPDNet-Light could be an option for environments with limited computational resources.

\begin{table*}[!htbp]
\centering
\newcommand{\cl}{35}
\resizebox{\linewidth}{!}
{
\begin{tabular}{c | c|c|c|c || c|c|c|c || c|c|c|c}
\toprule
\multirow{2}{*}{\bf Method} &\multicolumn{4}{c}{\bf Indoor} &\multicolumn{4}{c}{\bf Outdoor} &\multicolumn{4}{c}{\bf Combined}\\ \cline{2-13}
				
		& PSNR $\uparrow$ & SSIM $\uparrow$ & MAE $\downarrow$ & LPIPS $\downarrow$ & PSNR $\uparrow$ & SSIM $\uparrow$ & MAE $\downarrow$ & LPIPS $\downarrow$ & PSNR $\uparrow$ & SSIM $\uparrow$ & MAE $\downarrow$ & LPIPS $\downarrow$ \\ \midrule \midrule
		
{\bf DPDNet-Light}	 	&  27.08 &  0.824 & 0.030 &  0.225 & 22.81 & 0.701 & 0.053 & 0.309 &  24.89 & 0.761 & 0.042 & 0.268 \\ \midrule
{\bf DPDNet}	 	& {\bf 27.48} & {\bf 0.849} & {\bf 0.029} & {\bf 0.189} & {\bf 22.90} & {\bf 0.726} & {\bf 0.052} & {\bf 0.255} & {\bf 25.13} & {\bf 0.786} & {\bf 0.041} & {\bf 0.223} \\ \midrule
\end{tabular}
}
\caption{DPDNet with less blocks. The quantitative results of DPDNet-Light vs. our full DPDNet using four metrics. The testing on the dataset is divided into three scene categories: indoor, outdoor, and combined. The best results are in bold numbers. Our full DPDNet has the best results on all metrics for different categories. Nevertheless, DPDNet-Light can operate with less computational power and produce acceptable deblurring results. Note: the testing set consists of 37 indoor and 39 outdoor scenes.}
\label{tab:lessBlock}
\end{table*}

\subsection{DPDNet with different input sizes}\label{sec:inputSize}

Our DPDNet is a fully convolutional network.  This facilitates training with different input patch sizes with no change required in the network architecture. As such, we consider training with two different patch sizes, namely $256\times256$ pixels and $512\times512$ pixels referred to as DPDNet$_{256}$ and DPDNet$_{512}$, respectively.

Table~\ref{tab:inputSize} shows that the two different input sizes perform similarly. Particularly, input patch size does not change the performance drastically as long as it is larger than the blur size.

\begin{table*}[!htbp]
\centering
\newcommand{\cl}{35}
\resizebox{\linewidth}{!}
{
\begin{tabular}{c | c|c|c|c || c|c|c|c || c|c|c|c}
\toprule
\multirow{2}{*}{\bf Method} &\multicolumn{4}{c}{\bf Indoor} &\multicolumn{4}{c}{\bf Outdoor} &\multicolumn{4}{c}{\bf Combined}\\ \cline{2-13}
				
		& PSNR $\uparrow$ & SSIM $\uparrow$ & MAE $\downarrow$ & LPIPS $\downarrow$ & PSNR $\uparrow$ & SSIM $\uparrow$ & MAE $\downarrow$ & LPIPS $\downarrow$ & PSNR $\uparrow$ & SSIM $\uparrow$ & MAE $\downarrow$ & LPIPS $\downarrow$ \\ \midrule \midrule
		
{\bf DPDNet$_{256}$}	 	& 27.28 & 0.847 & {\bf 0.029} & 0.195 & 22.86 & {\bf 0.734} & {\bf 0.050} & 0.257 & 25.01 & {\bf 0.789} & {\bf 0.040} & 0.227 \\ \midrule
{\bf DPDNet$_{512}$}	 	&  {\bf 27.48} & {\bf 0.849} & {\bf 0.029} & {\bf 0.189} & {\bf 22.90} & 0.726 & 0.052 & {\bf 0.255} & {\bf 25.13} & 0.786 & 0.041 & {\bf 0.223} \\ \midrule
\end{tabular}
}
\caption{DPDNet with different input sizes. The quantitative results of DPDNet$_{256}$ vs. DPDNet$_{512}$ using four metrics. The testing on the dataset is divided into three scene categories: indoor, outdoor, and combined. The best results are in bold numbers. Both input sizes perform on par, in which the patch size does not change the performance drastically as long as it is larger than the blur size. Note: the testing set consists of 37 indoor and 39 outdoor scenes.}
\label{tab:inputSize}
\end{table*}

\subsection{DPDNet with different filtering ratios}\label{sec:filtering}

Homogeneous patches are inherently ambiguous in terms of incurred blur size, and do not provide useful information for network training~\cite{park2017unified}. As a result, filtering homogeneous patches can be beneficial to the trained network. In this section, different filtering ratios are examined including: $0\%$, $15\%$, $30\%$, and $45\%$; we refer to them as DPDNet$_{0\%}$, DPDNet$_{15\%}$, DPDNet$_{30\%}$, DPDNet$_{45\%}$, respectively.

In Table~\ref{tab:filtering}, we present the results of different filtering ratios. The $30\%$ filtering is a reasonable ratio that has the best quantitative results. Therefore, we filter $30\%$ of the extracted image patches based on the sharpness energy to train our proposed DPDNet as described in Sec.~6 of the main paper.

\begin{table*}[!htbp]
\centering
\newcommand{\cl}{35}
\resizebox{\linewidth}{!}
{
\begin{tabular}{c | c|c|c|c || c|c|c|c || c|c|c|c}
\toprule
\multirow{2}{*}{\bf Method} &\multicolumn{4}{c}{\bf Indoor} &\multicolumn{4}{c}{\bf Outdoor} &\multicolumn{4}{c}{\bf Combined}\\ \cline{2-13}
				
		& PSNR $\uparrow$ & SSIM $\uparrow$ & MAE $\downarrow$ & LPIPS $\downarrow$ & PSNR $\uparrow$ & SSIM $\uparrow$ & MAE $\downarrow$ & LPIPS $\downarrow$ & PSNR $\uparrow$ & SSIM $\uparrow$ & MAE $\downarrow$ & LPIPS $\downarrow$ \\ \midrule \midrule
{\bf DPDNet$_{0\%}$}	 	& 27.21 & 0.838 & 0.030 & 0.205 & 22.86 & 0.721 & {\bf 0.051} & 0.275 &  24.98 & 0.778 & {\bf 0.041} & 0.241 \\ \midrule
{\bf DPDNet$_{15\%}$}	 	&  27.19 & 0.840 & {\bf 0.029} & 0.194 & {\bf 22.94} & 0.721 & 0.052 & {\bf 0.254} &  25.01 & 0.779 & {\bf 0.041} & 0.225 \\ \midrule
{\bf DPDNet$_{30\%}$}	 	& {\bf 27.48} & {\bf 0.849} & {\bf 0.029} & {\bf 0.189} & 22.90 & {\bf 0.726} & 0.052 & 0.255 & {\bf 25.13} & {\bf 0.786} & {\bf 0.041} & {\bf 0.223} \\ \midrule
{\bf DPDNet$_{45\%}$}	 	&   27.21 &  0.839 & 0.030 &  0.194 & 22.90 & 0.724 & {\bf 0.051} & 0.258 &  25.00 & 0.780 & {\bf 0.041} & 0.227 \\ \midrule
\end{tabular}
}
\caption{DPDNet with different filtering ratios. The quantitative results of DPDNet$_{0\%}$ vs. DPDNet$_{15\%}$ vs. DPDNet$_{30\%}$ vs. DPDNet$_{45\%}$ using four metrics. The testing on the dataset is divided into three scene categories: indoor, outdoor, and combined. The best results are in bold numbers. The $30\%$ filtering is a reasonable ratio that has the best quantitative results and , thus, we pick it as a filtering ratio for our proposed framework. Note: the testing set consists of 37 indoor and 39 outdoor scenes.}
\label{tab:filtering}
\end{table*}

\subsection{DPDNet with different data types}\label{sec:dataType}

Our dataset provides high-quality images that are processed to an sRGB encoding with a lossless 16-bit depth per RGB channel.  Since we are targeting dual-pixel information which would be obtained directly in the camera's hardware, in a real hardware implementation we would expect to have such high bit-depth images.  However, since most standard encodings still rely on 8-bit image, we provide a comparison of training our DPDNet with 8-bit (DPDNet$_{8-bit}$) and 16-bit (DPDNet$_{16-bit}$) input data type.

Based on the numbers in Table~\ref{tab:dataType}, DPDNet$_{16-bit}$ has a slightly better performance. In particular, it has a lower LPIPS distance for all categories. As a result, training with 16-bit images is helpful due to the extra information embedded in, and is more representative of the hardware's data.

\newcommand{\figwid}{.496}
\begin{figure}[t]
  \newcommand{\figname}{1P0A1002}
  \centering

  \begin{subfigure}[t!]{\figwid\textwidth}
	\centering
	\includegraphics[width=\textwidth]{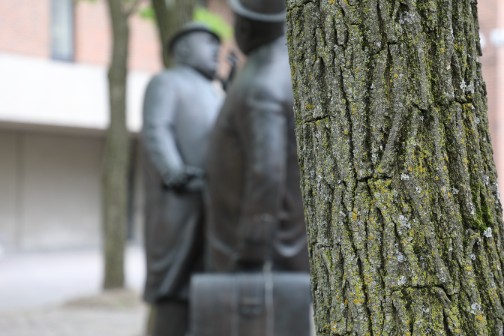}
	\caption{Blurred input image.}
  \end{subfigure}\hfill
  \begin{subfigure}[t!]{\figwid\textwidth}
	\centering
	\includegraphics[width=\textwidth]{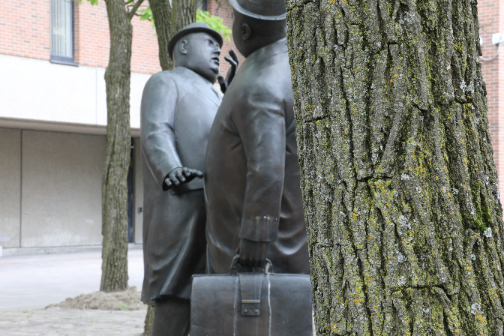}
	\caption{Ground truth sharp image.}
  \end{subfigure}

  \begin{subfigure}[t!]{\figwid\textwidth}
	\centering
	\includegraphics[width=\textwidth]{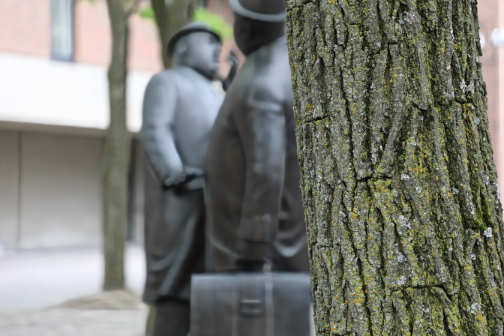}
	\caption{SRNet~\cite{tao2018scale} output image.}
  \end{subfigure}\hfill
  \begin{subfigure}[t!]{\figwid\textwidth}
	\centering
	\includegraphics[width=\textwidth]{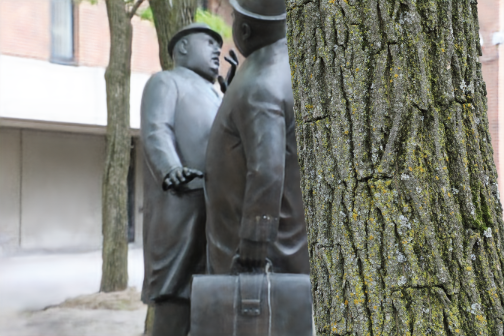}
	\caption{Our DPDNet output image.}
  \end{subfigure}
  \caption{Qualitative deblurring results using SRNet~\cite{tao2018scale} and our DPDNet.}\label{fig:qualResExMo}
\end{figure}

\begin{table}[!t]
\centering
\newcommand{\cl}{35}
\resizebox{\linewidth}{!}
{
\begin{tabular}{c | c|c|c|c || c|c|c|c || c|c|c|c}
\toprule
\multirow{2}{*}{\bf Method} &\multicolumn{4}{c}{\bf Indoor} &\multicolumn{4}{c}{\bf Outdoor} &\multicolumn{4}{c}{\bf Combined}\\ \cline{2-13}
				
		& PSNR $\uparrow$ & SSIM $\uparrow$ & MAE $\downarrow$ & LPIPS $\downarrow$ & PSNR $\uparrow$ & SSIM $\uparrow$ & MAE $\downarrow$ & LPIPS $\downarrow$ & PSNR $\uparrow$ & SSIM $\uparrow$ & MAE $\downarrow$ & LPIPS $\downarrow$ \\ \midrule \midrule
{\bf DPDNet$_{8-bit}$}	 	& 27.37 & 0.834 &  {\bf 0.029} & 0.196 & {\bf 23.10} & 0.723 & {\bf 0.052} & 0.258 &  {\bf 25.18} & 0.777 & {\bf 0.041} & 0.228 \\ \midrule
{\bf DPDNet$_{16-bit}$}	 	& {\bf 27.48} & {\bf 0.849} & {\bf 0.029} & {\bf 0.189} & 22.90 & {\bf 0.726} & {\bf 0.052} & {\bf 0.255} & 25.13 & {\bf 0.786} & {\bf 0.041} & {\bf 0.223} \\ \midrule
\end{tabular}
}
\caption{DPDNet with different data types. The quantitative results of DPDNet$_{8-bit}$ vs. DPDNet$_{16-bit}$ using four metrics. The testing on the dataset is divided into three scene categories: indoor, outdoor, and combined. The best results are in bold numbers. DPDNet$_{16-bit}$ has a slightly better performance, in which it has a lower LPIPS distance for all categories. Note: the testing set consists of 37 indoor and 39 outdoor scenes.}
\label{tab:dataType}
\end{table}


\begin{figure}[t]
\includegraphics[width=\linewidth]{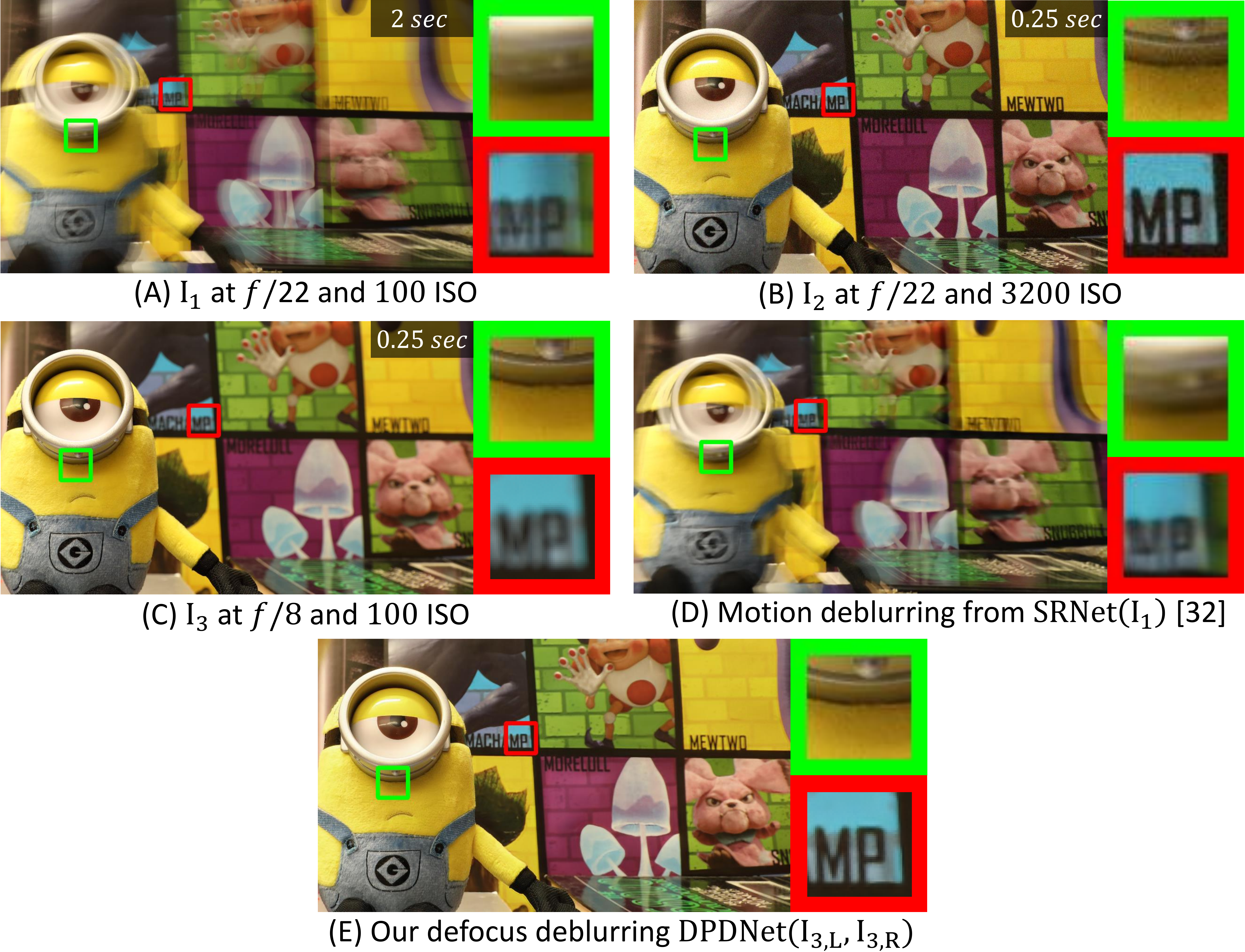}
\caption{Image noise, motion and defocus blur relation with a moving camera. The number shown on each image is the shutter speed. Zoomed-in cropped patches are also provided. (A) shows an image $\mathrm{I_1}$ suffers from motion blur. (B) shows an image $\mathrm{I_2}$ fixes the motion blur by increasing the ISO, however, $\mathrm{I_2}$ has more noise. (C) shows another image $\mathrm{I_3}$ handles the motion blur by increasing the aperture size, nevertheless, $\mathrm{I_3}$ suffers from defocus blur. (D) shows the results of deblurring $\mathrm{I_1}$ using the motion deblurring method SRNet~\cite{tao2018scale}. The image in (E) is the sharp and clean image obtained using our DPDNet to deblur $\mathrm{I_3}$.}\label{fig:useCase1}
\end{figure}

\section{Defocus and motion blur discussion}\label{sec:discussion}

One may be curious if motion blur methods can be used to address the defocus blur problem.  While defocus and motion blur both produce a blurring of the underlying latent image, the physical image formation process of these two types of blur are different.
Therefore, comparing with methods that solve for motion blur is not expected to give good results. However, for a validity check, we tested the scale recurrent motion deblurring method (SRNet) in~\cite{tao2018scale} using our testing set. This method achieved an average LPIPS of 0.452 and PSNR of 20.12, which is lower than all other existing methods that solve for defocus deblurring. Fig.~\ref{fig:qualResExMo} shows results of applying motion deblurring network SRNet~\cite{tao2018scale} to input image from our dataset.


\section{Use cases}\label{sec:useCases}

As discussed in Sec. 1 of the main paper, we described how defocus blur is related to the size of the aperture used at capture time.  The size of the aperture is often dictated by the desired exposure which is a factor of aperture, shutter speed, and ISO setting.  As a result, there is a trade-off between image noise (from ISO gain), motion blur (shutter speed), and defocus blur (aperture). This trade off is referred to as the exposure triangle. In this section, we show some common cases, where defocus deblurring is required. 

\noindent{\textbf{Moving camera.}}~Global motion blur is more likely to occur with the moving cameras like hand-held cameras ($\mathrm{I_1}$ in Fig.~\ref{fig:useCase1}-A). One way to handle motion blur is to set a fast shutter speed and this can be done by either increasing the image gain (i.e., ISO) or the aperture size. However, higher ISO can introduce noise as stated in~\cite{plotz2017benchmarking} (Fig.~\ref{fig:useCase1}-B), and wider aperture can introduce undesired defocus blur as shown in $\mathrm{I_3}$ (Fig.~\ref{fig:useCase1}-C). For such case, we offer two solutions: apply motion deblurring method SRNet~\cite{tao2018scale} on $\mathrm{I_1}$ (result shown in Fig.~\ref{fig:useCase1}-D) or apply our defocus deblurring method on $\mathrm{I_3}$ (result shown in Fig.~\ref{fig:useCase1}-E). Our defocus deblurring method is able to obtain sharper and cleaner image as demonstrated in Fig.~\ref{fig:useCase1}-E.

\begin{figure}[t]
\centering
\includegraphics[width=\linewidth]{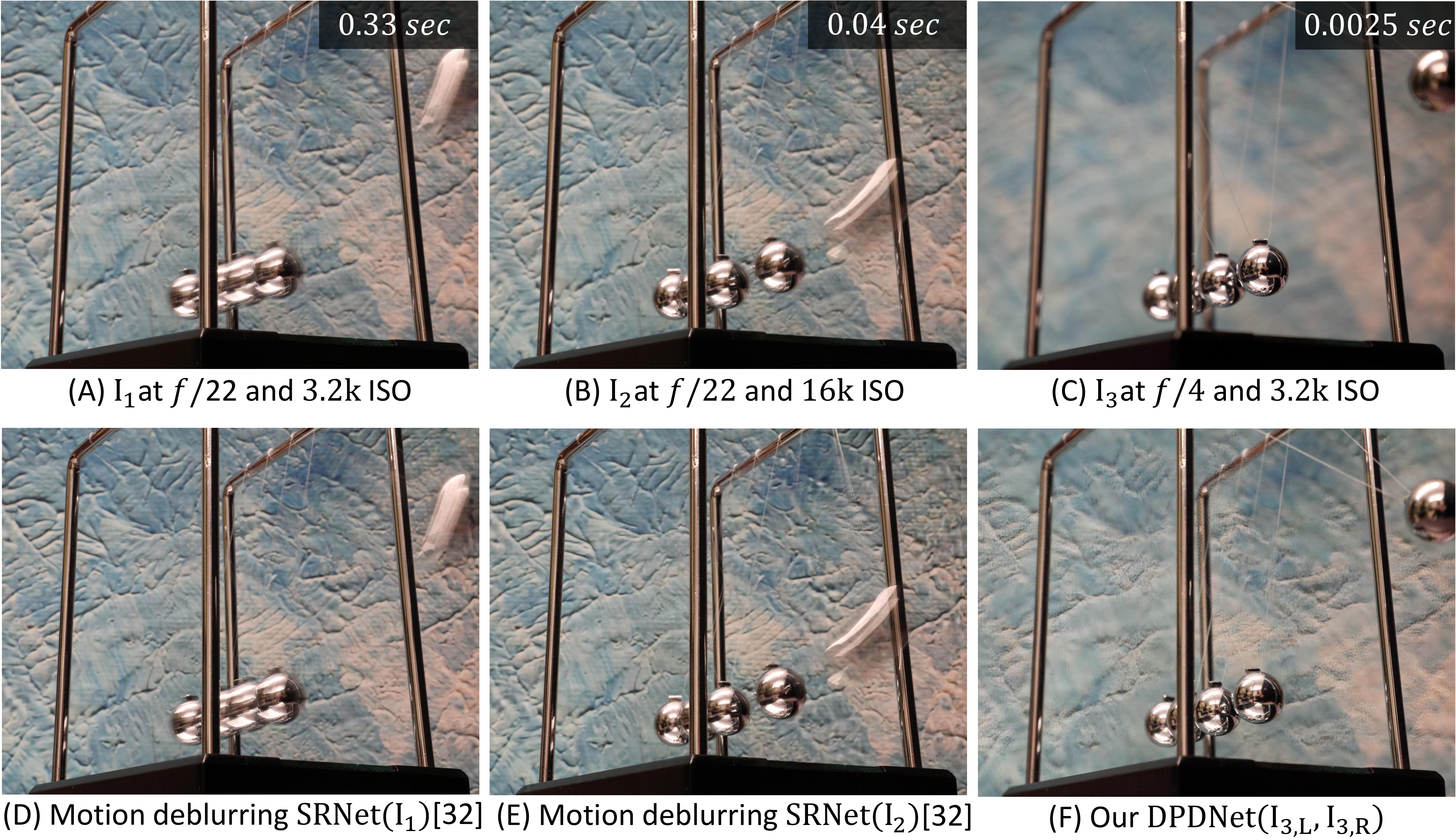}
\caption{Motion and defocus blur relation with a moving object. The number shown on each image is the shutter speed. (A) shows an image $\mathrm{I_1}$ has a moving object that suffers from motion blur. Image $\mathrm{I_2}$ in (B) tries to fix the motion blur by increasing the ISO, but the motion blur is still pronounced. $\mathrm{I_3}$ in (C) handles the motion blur by setting the aperture wide, nevertheless, it introduces defocus blur. (D) and (E) show the results of deblurring $\mathrm{I_1}$ and $\mathrm{I_2}$, respectively, using the motion deblurring method SRNet~\cite{tao2018scale}. The image in (F) is sharp and obtained by drblurring $\mathrm{I_3}$ using our DPDNet.}\label{fig:useCase2}
\end{figure}

\noindent{\textbf{Moving object.}}~In this scenario, we have a stationary camera, with a scene object that is moving (i.e., Newton's cradle in Fig.~\ref{fig:useCase2}). Fig.~\ref{fig:useCase2}-A shows an image with motion blur, in which the object speed is higher than the shutter speed. In Fig.~\ref{fig:useCase2}-B, the ISO is significantly increased in order to make the shutter speed faster, nevertheless, the pendulum speed remains the fastest and the motion blur is pronounced. Another way to increase the shutter speed is to open the aperture wider as shown in Fig.~\ref{fig:useCase2}-C and this setting handles the motion blur. However, capturing at wider aperture introduces the undesired defocus blur. To get a sharper image, we can use the motion deblurring method SRNet~\cite{tao2018scale} to deblur $\mathrm{I_1}$ (result shown in Fig.~\ref{fig:useCase2}-D) and $\mathrm{I_2}$ (result shown in Fig.~\ref{fig:useCase2}-E), or apply our defocus deblurring method on $\mathrm{I_3}$ (result shown in Fig.~\ref{fig:useCase2}-F). Our defocus deblurring method is able to obtain sharper image compared to motion deblurring method as demonstrated in Fig.~\ref{fig:useCase2}-F.

\section{DPDNet performance for a smartphone DP sensor}\label{sec:pixel}

\begin{figure}[t]
\centering
\includegraphics[width=\linewidth]{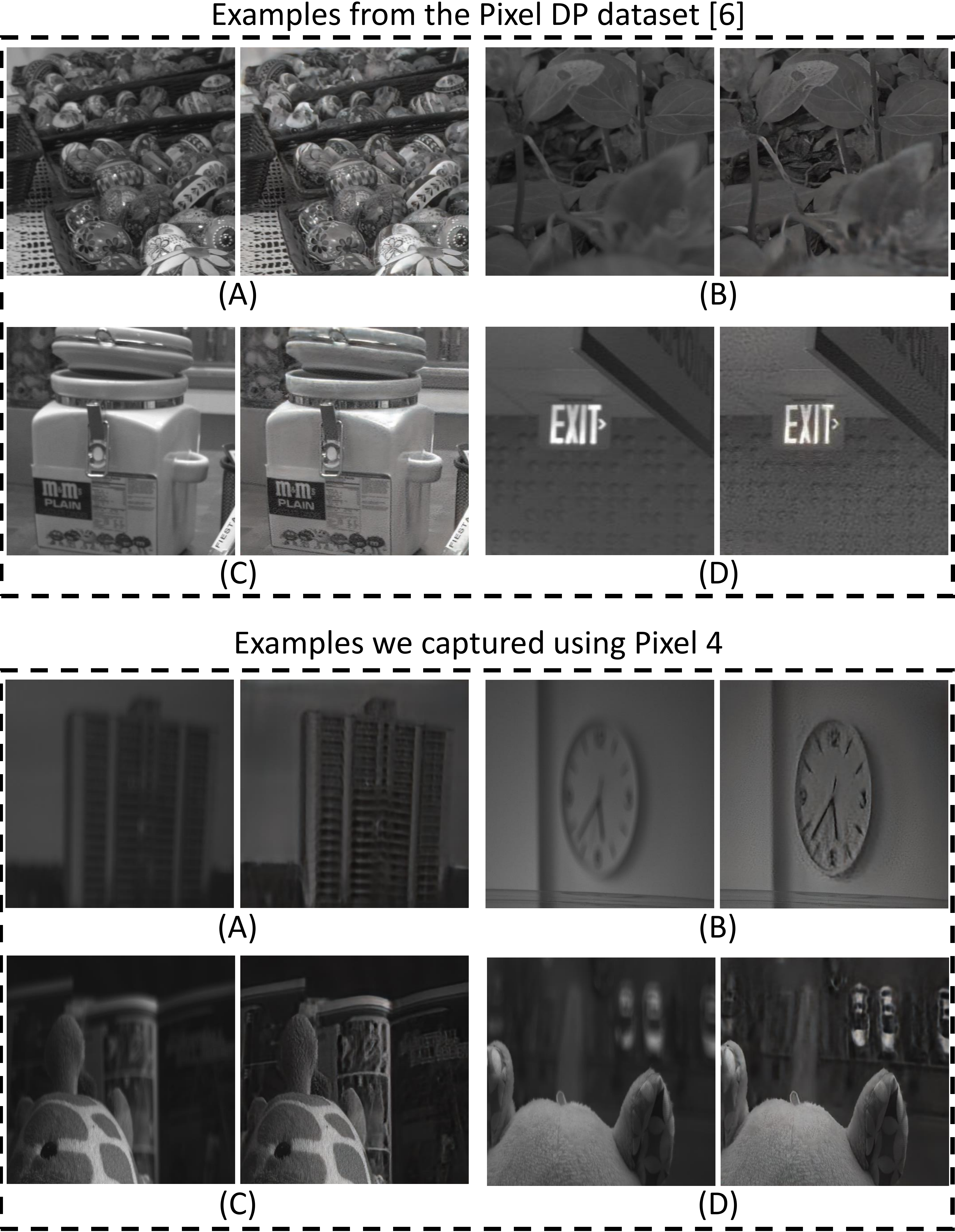}
\caption{The results of using our DPDNet to deblur images captured by Pixel smartphone camera. The image on the left is the combined input image with defocus blur and the one on the right is deblurred one. Our DPDNet is able to generalize well for images captured by a smartphone camera.}\label{fig:pixel}
\end{figure}

In this section, we test our DPDNet on images captured with a smartphone. As we mentioned in Sec. 4 of the main paper, there are two camera manufacturers that provide DP data, namely, Google Pixel 3 and 4 smartphones and Canon EOS 5D Mark IV DSLR.   The smartphone camera currently has limitations that make it challenging to train the DPDNet with.  First, the Google Pixel smartphone cameras do not have adjustable apertures, so we are unable to capture  corresponding ``sharp'' images using a small aperture as we did with the Canon camera.  Second, the data currently available from the Pixel smartphones are not full-frame, but are limited to only one of the Green channels in the raw-Bayer frame.  Finally, the smartphone has a very small aperture so most images do not exhibit defocus blur.  In fact, many smartphone cameras synthetically apply defocus blur to produce the shallow DoF effect.

As a result, the experiments here are provided to serve as a proof of concept that our method should generalize to other DP sensors.  To this end, we examined DP images available in the dataset from ~\cite{garg2019learning} to find images exhibiting defocus blur.  The L/R views of these images are available in the ``animated\_dp\_examples'' directory---located at the same directory as this pdf file.

To use our DPDNet, we replicate the single green  channel to be 3-channel image to match our DPDNet input. Fig.~\ref{fig:pixel} shows the deblurring results on images captured by Pixel camera. The image on the left is the input combined image and the image on the right is the deblurred one using our DPDNet. Note that the Pixel android application, used to extract DP data, does not provide the combined image~\cite{google2019api}. To obtain it, we average the two views. Fig.~\ref{fig:pixel} visually demonstrates that our DPDNet is able to generalize and deblur for images that are captured by the smartphone camera. Because it is not possible to  adjust aperture on the smartphone camera to capture a ground truth image, we cannot report quantitative numbers. The results of two more full images are shown in Fig.~\ref{fig:qualResExPixel1}.

\newcommand{\figcropPix}{1.6}
\begin{figure}[t]
  \centering

  \begin{subfigure}[t!]{\figwid\textwidth}
	\centering
	\includegraphics[trim={0 \figcropPix cm 0 \figcropPix cm},clip,width=\textwidth]{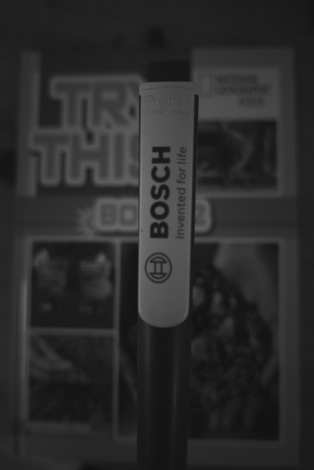}
	\caption{Blurred input image.}
  \end{subfigure}\hfill
  \begin{subfigure}[t!]{\figwid\textwidth}
	\centering
	\includegraphics[trim={0 \figcropPix cm 0 \figcropPix cm},clip,width=\textwidth]{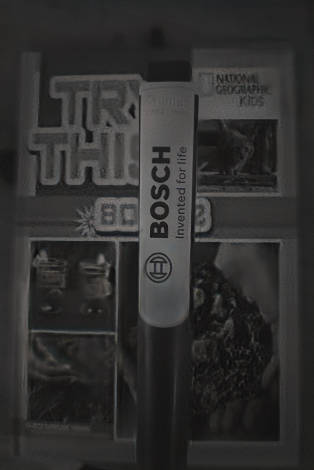}
	\caption{Our DPDNet output image.}
  \end{subfigure}

  \begin{subfigure}[t!]{\figwid\textwidth}
	\centering
	\includegraphics[trim={0 \figcropPix cm 0 \figcropPix cm},clip,width=\textwidth]{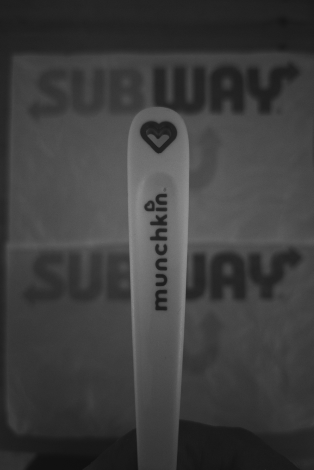}
	\caption{Blurred input image.}
  \end{subfigure}\hfill
  \begin{subfigure}[t!]{\figwid\textwidth}
	\centering
	\includegraphics[trim={0 \figcropPix cm 0 \figcropPix cm},clip,width=\textwidth]{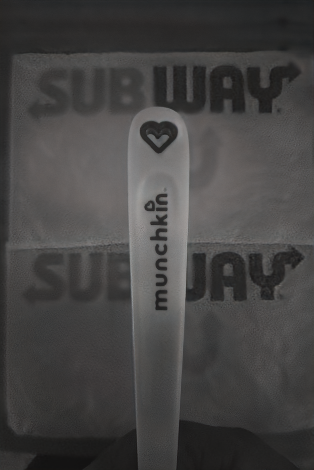}
	\caption{Our DPDNet output image.}
  \end{subfigure}
  \caption{Qualitative deblurring results using our DPDNet for images captured by a smartphone camera.}\label{fig:qualResExPixel1}
\end{figure}

\section{More results}\label{sec:qualRes}
\begin{table*}[t]
\centering
\newcommand{\cl}{35}
\resizebox{0.5\linewidth}{!}
{
\begin{tabular}{c | c|c}
\toprule
\multirow{2}{*}{\bf Method} &\multicolumn{2}{c}{\bf Average LPIPS $\downarrow$} \\ \cline{2-3}
				
									& DP L view & DP R view \\ \midrule \midrule
{\bf EBDB~\cite{karaali2017edge}}	&  0.342 &  0.337 \\ \midrule
{\bf DMENet~\cite{lee2019deep}}		&  0.355 &  0.353 \\ \midrule
{\bf JNB~\cite{shi2015just}}		&  0.322 &  0.313 \\ \midrule
{\bf Our DPDNet}	 					&\multicolumn{2}{c}{\bf 0.223} \\ \midrule
\end{tabular}
}
\caption{Average LPIPS evaluation of a single DP view separately.}
\label{tab:singleDpView}
\end{table*}
\noindent{\textbf{Quantitative results.}}~
In Table~\ref{tab:singleDpView}, we provide evaluation of other methods on a single DP view separately using the average LPIPS.
Note that a single DP L or R view is formed with a half-disc point spread function in the ideal case. When the two views are combined to form the final output image; the blur kernel would look like a full-disc kernel~\cite{punnappurath2020modeling}. Non-blind defocus deblurring methods assume full-disc kernel and the blur kernel of the combined image aligns more with their assumption. More details about DP view formation and modeling DP blur kernels can be found in~\cite{punnappurath2020modeling}.

\begin{table*}[t]
\centering
\newcommand{\cl}{35}
\resizebox{0.5\linewidth}{!}
{
\begin{tabular}{c | c }
\toprule
{\bf Method} &{\bf Average LPIPS  $\downarrow$} \\ \midrule
{\bf EBDB~\cite{karaali2017edge}}	&  0.229 \\ \midrule
{\bf DMENet~\cite{lee2019deep}}		&  0.216 \\ \midrule
{\bf JNB~\cite{shi2015just}}		&  0.207 \\ \midrule
{\bf Our DPDNet}	 					&{\bf 0.104}\\ \midrule
\end{tabular}
}
\caption{Average LPIPS evaluation of the images used to test DPDNet robustness to different aperture settings.}
\label{tab:robustnessAperture}
\end{table*}

In addition to above, we report in Table~\ref{tab:robustnessAperture} the average LPIPS numbers for other methods on the images used to test DPDNet robustness to different aperture settings. Note that the LPIPS numbers here are lower than numbers in Table 1 of the main paper. The reason is that for the robustness test we used f/10 and f/16, which results in less defocus blur compared to the images captured at f/4 (a much wider aperture than f/10 and f/16).

\end{document}